\documentclass{aa}

\usepackage{natb2e}
\renewcommand\citep[1]{(\citealt{#1})}  
\newcommand\citepf[1]{(\citealt*{#1})}    

\usepackage{epsfcopy}

\newcommand\zzz[2]{#2}  

\def\hMpc{\mbox{$h^{-1}\mbox{\rm Mpc}$}}

\def\centreline{\centerline}
\def\gtapprox{\,\lower.6ex\hbox{$\buildrel >\over \sim$} \, }
\def\ltapprox{\,\lower.6ex\hbox{$\buildrel <\over \sim$} \, }
\def\sun{\odot}

\def\e{ {\scriptstyle \times} 10^}

\def\arcs{\ifmmode {'' }\else $'' $\fi}     
\def\arcm{\ifmmode {' }\else $' $\fi}     
\def\deg{\ifmmode^\circ\else$^\circ$\fi}    

\def\frtoday{le\space\number\day\space\ifcase\month\or
  janvier\or f\'evrier\or mars\or avril\or mai\or juin\or
  juillet\or ao\^ut\or septembre\or octobre\or novembre\or d\'ecembre\fi\space \number\year}

\def\nlev_tiny{n_{\mbox{\rm \tiny levels}}} 

\def\tdyn{t_{\mbox{\rm \small dyn}}}

\def\Mcold{M_{\mbox{\rm \small cold}}}

\def\pmed{P_{\mbox{\rm \small med}}}
\def\pmin{P_{\mbox{\rm \small min}}}
\def\Nmin{N_{\mbox{\rm \small min}}}

\def\arfus{{\sc ArFus}}
\def\SS{Sect.~}

\newcommand\joref[5]{#1, #5, {#2, }{#3, } #4}

\newcommand\epref[3]{#1, #3, #2}

\def\MNRAS{MNRAS}
\def\apj{ApJ}                 
\def\apjs{ApJS}                 
\def\aanda{A\&A}            
\begin{document}


\title{Star formation 
losses due to tidal debris in `hierarchical' galaxy formation}

\def\IAP{Institut d'Astrophysique de Paris, 98bis Bd Arago, F-75.014 Paris,
France}
\def\Stras{Observatoire astronomique de Strasbourg, CNRS, 
11, rue de l'Universit\'e, F-67.000 Strasbourg, France}
\def\CAMK{Nicolaus Copernicus Astronomical Center, 
ul. Bartycka 18, 00-716 Warsaw, Poland}
\def\IUCAA{Inter-University Centre for Astronomy and Astrophysics, 
 Post Bag 4, Ganeshkhind, Pune, 411 007, India}
\def\DAEC{DAEC, Observatoire de Paris-Meudon, 5, place Jules Janssen, 
F-92.195, Meudon Cedex, France}



\author{B. F. Roukema\inst{1,2,3,4} \and S. Ninin\inst{1} \and
J. Devriendt\inst{1} \and F. R. Bouchet\inst{1} \and B. Guiderdoni\inst{1}
\and  G. A. Mamon\inst{1,5}} 

\offprints{B. F. Roukema, present address 
DARC/LUTH, Observatoire de Paris-Meudon, 5, place Jules Janssen, 
F-92.195, Meudon Cedex, France}

\institute{ {\IAP}
(boud.roukema@obspm.fr, ninin@iap.fr, devriend@iap.fr, 
bouchet@iap.fr, guider@iap.fr, gam@iap.fr) 
\and 
 {\Stras} \and {\CAMK} \and {\IUCAA} \and {\DAEC} }

\date{\frtoday}

\authorrunning{B. F. Roukema et al.}
\titlerunning{Tidal debris losses in hierarchical galaxy formation}

\abstract{
$N$-body studies have previously shown that the bottom-up hierarchical 
formation of dark matter haloes is not as monotonic as implicitly assumed
in the Press-Schechter formalism. During and following halo mergers, 
matter can be ejected into tidal tails, shells or low density `atmospheres'
outside of the successor haloes' virialisation radii (or group-finder
outermost radii). The implications that 
the possible truncation of star formation in these tidal
`debris' may have for observational galaxy statistics 
are examined here using the {\arfus} $N$-body plus semi-analytical 
galaxy modelling software for standard star formation hypotheses.
\\
In the $N$-body simulations studied, the debris
typically remain close to the successor halo and fall back
into the successor haloes given sufficient time. A maximum debris loss
of around 16\% is found for redshift intervals of around $\Delta
z=0.4$ at $z\sim 1,$ with little dependence on the matter density
parameter $\Omega_0$ and the cosmological constant $\lambda_0$.
\\
Upper and lower bounds on stellar losses implied
by a given set of $N$-body simulation output data can be investigated
by choice of the merging/identity criterion of haloes between
successive $N$-body simulation output times.  A {\em median}
merging/identity criterion is defined and used to deduce an upper
estimate of possible star formation and stellar population losses.  A
{\em largest successor} merging/identity criterion is defined to
deduce an estimate which minimises stellar losses.
\\
The losses for star formation and luminosity functions 
are strongest for low luminosity galaxies --- a likely consequence of
the fact that the debris fraction is highest for low mass haloes ---
and at intermediate redshifts $(1\ltapprox z \ltapprox 3)$. The losses
in both cases are mostly around 10\%-30\%, have some dependence on
$\Omega_0$ and negligible dependence on $\lambda_0.$ This upper bound
on likely losses in star formation rates and stellar populations is
smaller than the uncertainties in estimates of corresponding
observational parameters. Hence, it may not be urgent to include a
correction for this in Press-Schechter based galaxy formation models,
except when statistics regarding dwarf galaxies are under study.
}

\maketitle


\def\ffrac{
\begin{figure}
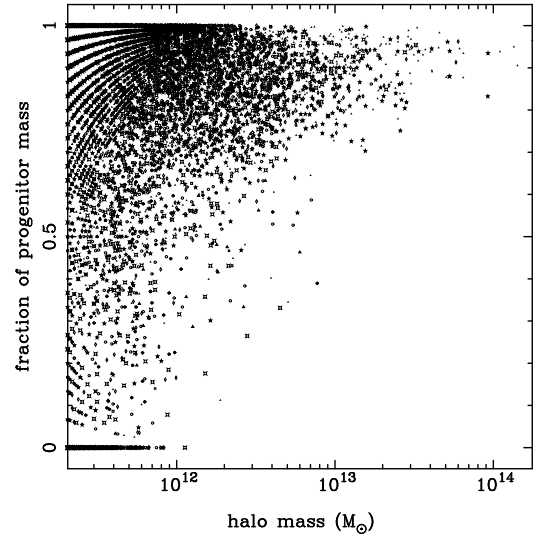

\centering 
 \centreline{\epsfxsize=7cm
\zzz{ \epsfbox[56 38 464 462]{"`gunzip -c frac.ps.gz"}}
  {\epsfbox[56 38 464 462]{"M9599f01.eps"}}
}
\caption[]{The fraction $P$ of a progenitor halo present in 
a successor halo between two $N$-body simulation output times separated
by $\Delta z=0.5,$ for successive output time pairs 
from $z=6$ to $z=0,$ for the
$(\Omega_0=1,\lambda_0=0)$ CDM-like simulation detailed in \SS\ref{s-nbody}, 
as a function of progenitor halo mass. 
Different symbols indicate different output time pairs, but to 
avoid crowding are too small to be easily distinguishable 
(the `squares with pointy corners' representing the interval
$ z \in [2.5,2.0]$ may be distinguishable, 
and possibly also the solid triangles for 
$ z \in [0.5,0.0]$).
The value of $P$ is unity
for the assumption of monotonic, spherical infall.
Note that the effects of discreteness 
show up as patterns at the left-hand extreme
of the figure, even though a minimum of 30 particles 
is required per halo. The points with $P=0$ represent haloes for which
no successor at all exists in the output time following by $\Delta z=0.5$.

\label{f-frac} 
}
\end{figure}
} 

\def\fpmeddelz{
\begin{figure}
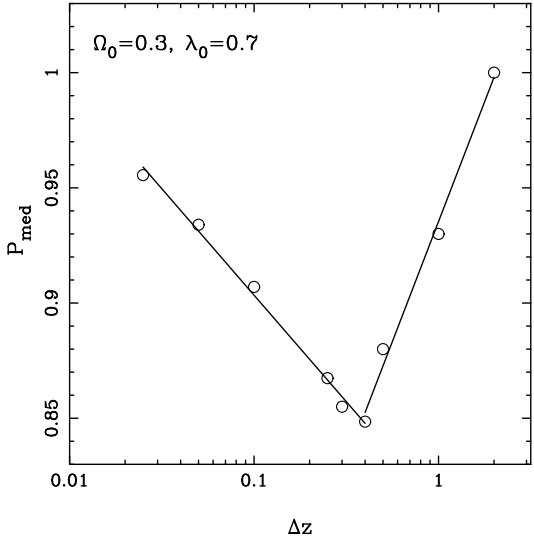

\centering 
 \centreline{\epsfxsize=7cm
\zzz{ \epsfbox[48 34 459 452]{"`gunzip -c pmeddelz.ps.gz"}}
  {\epsfbox[48 34 459 452]{"M9599f02.eps"}}
}
\caption[]{Dependence of the median fraction 
${\pmed}$ of a progenitor halo present in 
a successor halo as a function of $N$-body simulation output time
redshift interval $\Delta z,$ for the 
$(\Omega_0=0.3,\lambda_0=0.7)$ 
CDM-like simulation detailed in \SS\ref{s-nbody}, in the redshift
range $1 \ge z \ge 0.5$ for $\Delta z < 0.5$ and 
$6 \ge z \ge 0$ for $\Delta z \ge 0.5$. Each value
obtained from a set of simulation output times is shown as
a circle.
Linear fits to ${\pmed}$ as a function of
$\log_{10}(\Delta z)$ above and below the $\Delta z =0.4$ point
are shown here and given in eq.~(\protect\ref{e-pmedian}).
\label{f-pmeddelz} 
}
\end{figure}
} 

\def\fmedian{
\begin{figure}
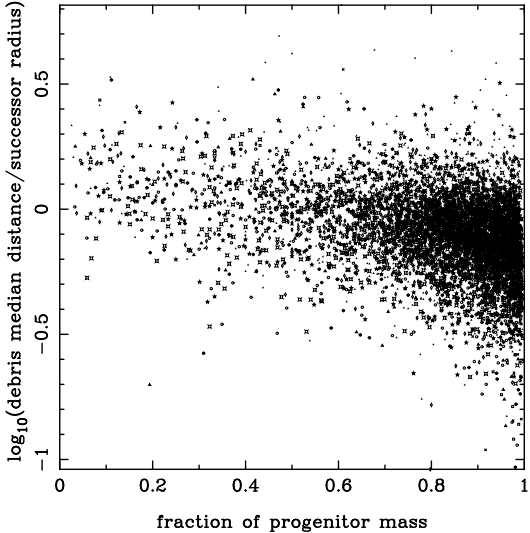

\centering 
 \centreline{\epsfxsize=7cm
\zzz{ \epsfbox[56 38 464 462]{"`gunzip -c median.ps.gz"}}
  {\epsfbox[56 38 464 462]{"M9599f03.eps"}}
}
\caption[]{The ratio of the median distance of debris to
the outermost radius of a successor halo, as a function of 
the fraction of a progenitor present in the successor,
for the simulation used in Fig.~\protect\ref{f-frac}. 
Symbols are as for Fig.~\protect\ref{f-frac}. 
\label{f-rmedian} 
}
\end{figure}
} 

\def\fttail{
\begin{figure}
{
\centering 
 \centreline{\epsfxsize=8cm
\zzz{ \epsfbox[56 40 552 555]{"`gunzip -c revirA.ps.gz"}}
  {\epsfbox[56 40 552 555]{"M9599f06.eps"}}
}
}
\caption[]{Example of debris in a $2.6${\hMpc} cube around an
interesting halo, shown in this figure and in Figs~\protect\ref{f-ttailB}
and \protect\ref{f-ttailC}.
Dark matter which, at $z=1.0$,
is either 
detected in a (primary) successor or newly formed 
halo (small black points, 
`virialised' matter),
detected in a secondary successor `halo' (large black points, i.e. 
`re-virialised' debris), or which constitutes ordinary debris 
(large gray points) is shown at the pair of output times 
used to determine these criteria: $z=1.5$ is the first
output time (this figure), 
$z=1.0$ is the second output time (Fig.~\protect\ref{f-ttailB}).
Background particles are not shown.
The point styles/shadings of any given particle 
are physically determined at $z=1.0$
and kept identical for the earlier image at $z=1.5$ and for the
later image at $z=0.5$.
Axes are labelled in comoving Mpc. Projections are rectilinear
$X-Y-Z$ as indicated. 
In this image ($z=1.5$), 
all the particles destined to become debris, shown
by large dots, clearly occupy the dense bodies of the two big
haloes. Many particles destined to fall into the 
haloes, i.e. destined to satisfy the group-finder criterion
at the following output time, are also visible as scatterings
of small dots around the haloes.
\label{f-ttail} 
}
\end{figure}
} 

\def\fttailB{
\begin{figure}
\centering 
 \centreline{\epsfxsize=8cm
\zzz{ \epsfbox[56 40 552 555]{"`gunzip -c revirB.ps.gz"}}
  {\epsfbox[56 40 552 555]{"M9599f07.eps"}}
}
\caption[]{As for Fig.~\protect\ref{f-ttail}, for the 
output time at which classification into the three 
particle types shown is carried out 
(background particles are ignored), i.e. at $z=1.0$.
Two small `re-virialised' shells are visible in the $X-Y$ image 
around the main halo: at the bottom-left and at the top of this
halo, i.e. mainly separated in the $Y$ direction, with a small
separation in the $X$ direction. A possible shell/tail is visible 
for the higher $X$ value main halo. The former halo is moving towards
greater $X$ values and will merge with the latter,
as it is falling along a large filament towards a deep potential
well (not shown here).
\label{f-ttailB} 
}
\end{figure}
} 

\def\fttailC{
\begin{figure}
\centering 
 \centreline{\epsfxsize=8cm
\zzz{ \epsfbox[56 40 552 555]{"`gunzip -c revirC.ps.gz"}}
  {\epsfbox[56 40 552 555]{"M9599f08.eps"}}
}
\caption[]{As for Figs~\protect\ref{f-ttail} and 
\protect\ref{f-ttailB}, for a later
output time, i.e. at $z=0.5$.
Both the `re-virialised' shells (large black dots) 
visible in the $X-Y$ view of the halo
at $z=1.0$ in Fig.~\protect\ref{f-ttailB} and the ordinary debris
(large gray dots) 
around the same halo have clearly fallen in to the main body of the
halo, which has moved further towards greater $X$ values to start
merging with its neighbour.
\label{f-ttailC} 
}
\end{figure}
} 

\def\fangsp{
\begin{figure}
\centering 
 \centreline{\epsfxsize=6cm
\zzz{ \epsfbox[52 36 460 452]{"`gunzip -c angsp.ps.gz"}}
  {\epsfbox[52 36 460 452]{"M9599f04.eps"}}
}
\caption[]{Distributions of angles $\theta$ (in degrees) between 
(i) the vector from the centre of a successor halo to either a halo
(non-debris) particle or a debris particle and 
(ii) the major axis of the successor halo,
for the $z\in[1.5,1.0]$ interval of 
the $(\Omega_0=0.3,\lambda_0=0.7)$ simulation 
(Sect.~\protect\ref{s-rmedian}). The thin curve is
for non-debris particles, the thick curve is for debris particles.
The non-debris vectors are clearly more aligned with the major
An isotropic distribution would follow a $\sin\theta$ function.
\label{f-angsp} 
}
\end{figure}
} 

\def\fangve{
\begin{figure}
\centering 
 \centreline{\epsfxsize=6cm
\zzz{ \epsfbox[52 36 460 452]{"`gunzip -c angve.ps.gz"}}
  {\epsfbox[52 36 460 452]{"M9599f05.eps"}}
}
\caption[]{As for Fig.~\protect\ref{f-angsp}, except that
distributions of angles $\theta_{\mbox{v}}$  (in degrees) 
between debris and non-debris 
particles' {\em velocity} vectors
and successor halo (spatial) major axes, using the successor halo 
velocity and spatial centres
as origins, are shown. As for Fig.~\protect\ref{f-angsp}, 
the thin curve is for non-debris particles, the thick curve is
for debris particles. The two distributions are not as strikingly
different as those in Fig.~\protect\ref{f-angsp}.
\label{f-angve} 
}
\end{figure}
} 

\def\fmfun{
\begin{figure}
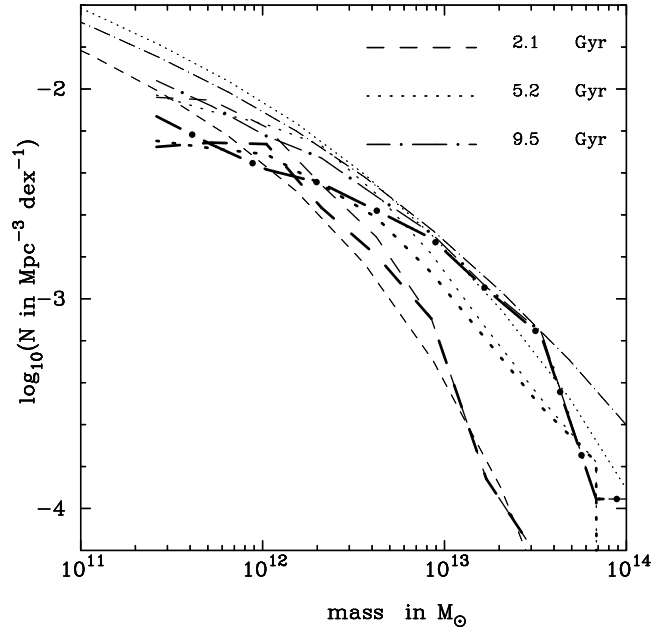

\centering 
 \centreline{\epsfxsize=8.5cm
\zzz{ \epsfbox[43 29 549 523]{"`gunzip -c mf.ps.gz"}}
  {\epsfbox[43 29 549 523]{"M9599f09.eps"}}
}
\caption[]{Halo mass function evolution (thick lines) 
based on the merger history 
tree interpolated between the $N$-body simulation outputs  
referred to in Fig.~\protect\ref{f-frac}. The very thin curves
indicate Press-Schechter mass functions for 
the curvature and perturbation spectrum parameters indicated
above, where a top-hat window function, $\delta_{c0}=1.3$ and 
a short wavelength smooth truncation at 1/64 of the box size 
are applied. The medium thickness curves are for the 
{\em largest successor} merger/identity criterion 
(\SS\protect\ref{s-largest}),
i.e. including debris, 
and the very thick curves are for the  
{\em median} merger/identity criterion 
(\SS\protect\ref{s-median}),
where $P\equiv 2{\pmed} - 1 = 78\%,$
i.e. excluding a significant
fraction of debris. The {\em median} criterion mass functions are slightly
lower than the {\em largest successor} criterion mass functions at
the low mass end.
\label{f-mfun} 
}
\end{figure}
} 

\def\fmadau{
\begin{figure}
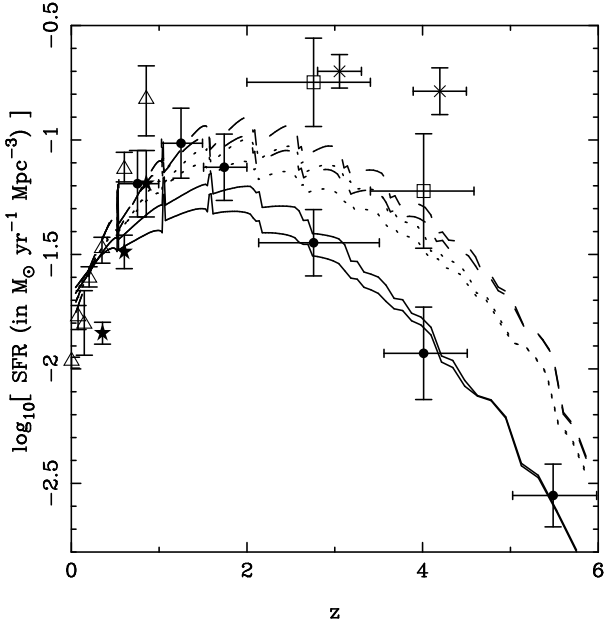

\centering 
 \centreline{\epsfxsize=8cm
\zzz{ \epsfbox[51 43 462 455]{"`gunzip -c madau.ps.gz"}}
  {\epsfbox[51 43 462 455]{"M9599f10.eps"}}
}
\caption[]{Global star formation rate histories for the `largest successor' 
merger/identity criterion (upper curve of each pair) 
and the `median' criterion (lower
curve of each pair). The solid, dashed and dotted curves are for the 
models with $(\Omega_0, \lambda_0)=(1.0,0.0),$
$(0.3,0.7)$ and $(0.3,0.0)$ respectively.
Observational points [taken from summary in fig.~9 of 
\protect\citet{Blain99}] are shown as open triangles
\protect\citep{Gall96,Gron99,Trey98,TM98,HammF99},
filled stars \protect\citep{Lilly96},
filled circles \protect\citepf{Conn97,Mad96}, 
open squares (dust-corrected, 
\protect\citealt{Pett98a}) 
and `x' crosses (dust-corrected, 
\protect\citealt{Steid99}). All models and observational estimates are
converted to $(\Omega_0, \lambda_0,h)=(1.0,0.0,0.65)$ for plotting purposes.

\label{f-madau} 
}
\end{figure}
} 

\def\flfun{
\begin{figure}
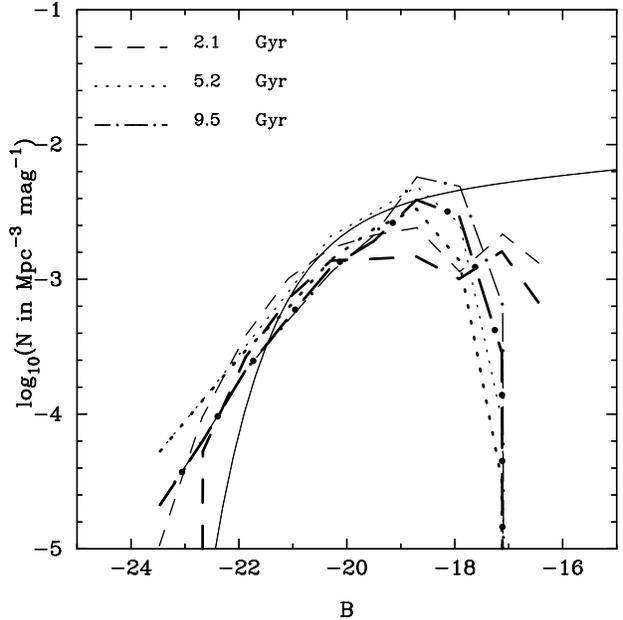

\centering 
 \centreline{\epsfxsize=9cm
\zzz{ \epsfbox[23 21 565 529]{"`gunzip -c lf.ps.gz"}}
  {\epsfbox[23 21 565 529]{"M9599f11.eps"}}
}
\caption[]{Luminosity function evolution, as for 
Fig.~\protect\ref{f-mfun}, medium thickness curves
 for exclusion of debris,
very thick curves for inclusion of debris.
The smooth curve to guide
the eye is a \protect\citet{Love92} parametrised observational 
luminosity function.
\label{f-lfun} 
}
\end{figure}
} 

\def\fflowchart{
\begin{figure}
\centerline{ \epsfxsize=8cm
\zzz{\epsfbox[0 0 503 610]{"`gunzip -c flowch.ps.gz"}}
{\epsfbox[0 0 503 610]{"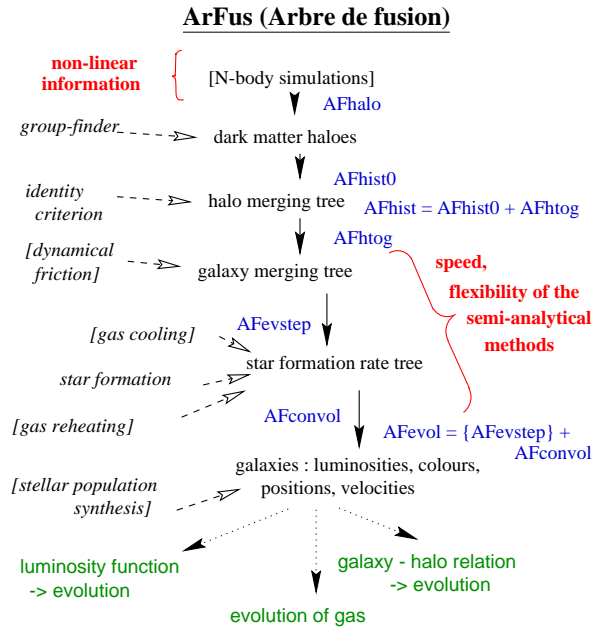"}}
}
\caption[]{Schematic flowchart of 
physical (in italics) and computational (in roman font) 
descriptions of the elements of the {\arfus} package. 
Key programme and subroutine names start with the prefix `AF'.
Square brackets indicate standard, non-original contributions. 
Sans serif fonts (bottom) indicate examples of numerical output.
\label{f-flowchart}
}
\end{figure}
} 

\def\ffhtogA{
\begin{figure}
\centerline{ \epsfxsize=8cm
\zzz{\epsfbox[0 0 506 478]{"`gunzip -c f_htogA.ps.gz"}}
{\epsfbox[0 0 506 478]{"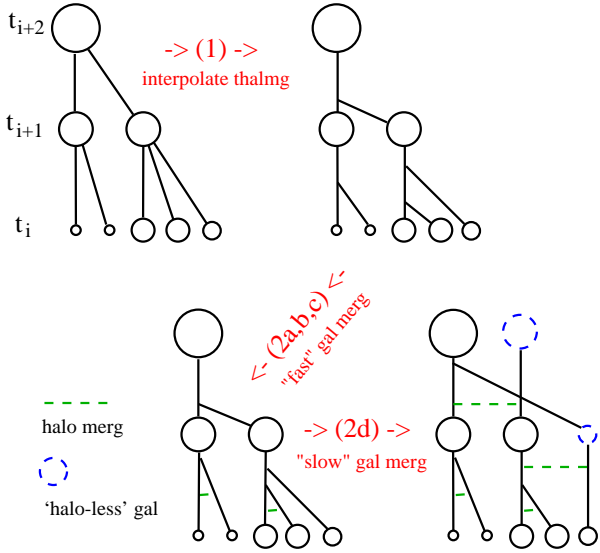"}}
}
\caption[]{Schematic description of the conversion of a
halo merging history tree into a galaxy merging history tree. 
Circles represent haloes and/or galaxies, line segments represent
identity and/or merging and time increases upwards.
See \protect\ref{s-afhtog} (1) to (2d) for details.
\label{f-f_htogA}
}
\end{figure}
} 

\def\ffhtogB{
\begin{figure}
\centerline{ \epsfxsize=8cm
\zzz{\epsfbox[0 0 495 475]{"`gunzip -c f_htogB.ps.gz"}}
{\epsfbox[0 0 495 475]{"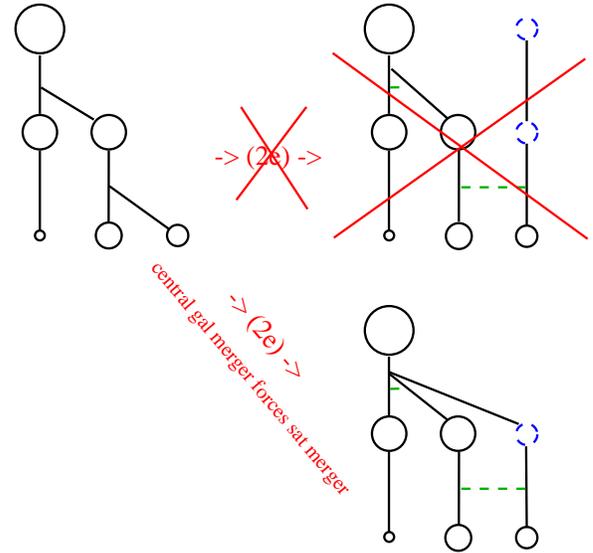"}}
}
\caption[]{Exceptional case requiring treatment by the algorithm 
which converts a 
halo merging history tree into a galaxy merging history tree. 
Symbols are as for Fig.~\protect\ref{f-f_htogA}.
See \protect\ref{s-afhtog} (2e) for details.
\label{f-f_htogB}
}
\end{figure}
} 

\def\tmidcrit{
\begin{table}
\caption{Summary of the different 
merging/identity criterion choices, where ${\pmin}$ is the minimum
fraction of particles of a halo at time $t_i$ present in a halo
at time $t_{i+1}$ in order for a merging/identity link to be defined
between the two haloes, and $N$ is the number of particles in the
halo at $t_{i+1}$. The $50\%$ criterion of \protect\citet{KCDW99a} differs
slightly from the $50\%$ criterion of \protect\citet{RPQR97}, and allows
recuperation of `lost' galaxies at later time steps.
\label{t-midcrit}}
$$\begin{array}{c c c  } \hline 
\pmin & \mbox{name} & \mbox{comments} \\
0.5 & \mbox{50\%} & \mbox{\protect\citet{RPQR97}} \\
0.5 & \mbox{--} & \mbox{\protect\citet{KCDW99a}} \\
2 \pmed -1 & \mbox{median} & \mbox{loss of debris; this paper} \\
1/N & \mbox{largest successor} & \mbox{monotonic infall; this paper} \\
\hline
\end{array}$$
\end{table}
}  

\def\tmedian{
\begin{table}
\caption{Median merging fractions $P$ as a function of redshift 
$z_i$ and of 
redshift interval $\Delta z$ 
between $N$-body simulation output times at $z_i$ and $z_i -\Delta z$,
for the simulations of metric parameters indicated. Note that 
although the merging fractions for the two low density models are
identical to two decimal places, several values differ at the third
decimal place (not quoted).
\label{t-median}}
$$\begin{array}{l ccc c c} \hline 
%
 (\Omega_0, \lambda_0)= &(1.0,0.0) &
(0.3,0.7) &(0.3,0.7) &(0.3,0.7) 
& (0.3,0.0) \\
z_i & \Delta z=0.5   & \Delta z=0.5 & \Delta z =1 & \Delta z =2 & \Delta z=0.5
 \\ \hline
%
%
6   & 1.00&        1.00 & 1.00 & 1.00 &1.00\\ 
5.5 & 1.00&        1.00 & &&1.00\\ 
5   & 1.00&        1.00 & 1.00 &&  1.00\\ 
4.5 & 1.00&         0.98 & &&0.98\\ 
4   & 0.99&        0.97 & 0.96 & 1.00 &0.97\\  
3.5 & 0.97&         0.94 & &&0.94\\ 
3   & 0.93&         0.90 & 0.89 &&  0.90\\ 
2.5 & 0.90&         0.87 & &&0.87 \\ 
2   & 0.86&         0.86 & 0.90 & 1.00 &0.86\\  
1.5 & 0.86&         0.85 & &&0.85\\ 
1   & 0.84&         0.85 & 0.95 &&  0.85\\ 
0.5 & 0.88&         0.88 & &&0.88\\ 
\hline
\multicolumn{4}{l}{\mbox{overall medians}} \\
    & 0.89&  0.88 & 0.93 & 1.00 &0.88\\
\hline
\end{array}$$
\end{table}
}  

\def\tmedianPoiss{
\begin{table}
\caption{Dependence on minimum number of particles per halo of
median merging fractions $P$, as a function of redshift $z_i$
between $N$-body simulation output times at $z_i$ and $z_i -0.5$,
for the $(\Omega_0, \lambda_0)= (0.3,0.7)$ simulation.
(The column for $\Nmin=30$ is identical to that in
Table~\protect\ref{t-median}.)
\label{t-medianPoiss}}
$$\begin{array}{l ccc c c} \hline 
z_i &  \Nmin=10  & \Nmin=30 & \Nmin=100 
\rule[-1.3ex]{0ex}{1ex}
 \\ \hline
6   & 1.00 & 1.00 &  0.99 \\
5.5 & 1.00 & 1.00 &  0.99 \\
5   & 1.00 & 1.00 &  0.99  \\
4.5 & 1.00 & 0.98 &  0.97 \\
4   & 0.95 & 0.97 &  0.95 \\
3.5 & 0.90 & 0.94 &  0.93 \\
3   & 0.82 & 0.90 &  0.92 \\
2.5 & 0.73 & 0.87 &  0.91 \\
2   & 0.67 & 0.86 &  0.90 \\
1.5 & 0.67 & 0.85 &  0.88 \\
1   & 0.70 & 0.85 &  0.87 \\
0.5 & 0.80 & 0.88 &  0.87 \\
\hline
\multicolumn{4}{l}{\mbox{overall medians}} \\
    & 0.79 &   0.88 & 0.89 \\
\hline
\end{array}$$
\end{table}
}  

\def\tmedianbis{
\begin{table}
\caption{Median merging fractions $P$ as a function of redshift $z_i$ and of 
redshift interval $\Delta z$, for smaller redshift intervals 
($z_i$, $z_i -\Delta z$) than those of
Table~\protect\ref{t-median}, for the simulation with 
$(\Omega_0=1, \lambda_0) = (0.3,0.7)$.
\label{t-medianbis}}
$$\begin{array}{l ccc c c} \hline 
 z_i 
& \Delta z=0.025& \Delta z=0.05 & \Delta z = 0.1 & \Delta z =0.25  
 \\ \hline

1.0   &&       0.94 & 0.91 & 0.86 \\ 
0.95  &&       0.94 &        \\ 
0.9   & 0.96 & 0.93 & 0.91&    \\ 
0.875 & 0.96 &&&          \\ 
0.85  & 0.96 & 0.94 &\\
0.825 & 0.96 \\
0.8   & 0.95 & 0.93 & 0.91 \\
0.775 & 0.96 \\
0.75  & 0.95 & 0.93 &      & 0.87  \\
0.725 & 0.96\\
0.7   & 0.95 & 0.93 & 0.90 \\
0.675 & 0.95 \\
0.65  & 0.95 & 0.93 \\
0.625 & 0.96 \\
0.6   & 0.96 & 0.93 & 0.91 \\
0.575 & 0.96 \\
0.55  & 0.95 & 0.93 \\
0.525 & 0.95 \\
\hline
\multicolumn{4}{l}{\mbox{overall medians}} \\
    & 0.96 & 0.93 & 0.91 &  0.87 \\
\hline
\end{array}$$
\end{table}
}  

\def\tangsp{
\begin{table}
\caption{Mean angles $\theta$ (defined in caption of 
Fig.~\protect\ref{f-angsp})
for non-debris halo (H) and debris (D) particles, their 
difference $\delta\theta$, 
and the $1\sigma$ uncertainty in the difference 
$\Delta(\delta \theta)$ 
[the uncertainty 
$\Delta(\delta\theta)$ is estimated by using the standard
deviations of $\theta$, taking standard errors in the
mean, and assuming Gaussian error propagation],
as a function of redshift $z_i$, 
between $N$-body simulation output times at $z_i$ and $z_i -0.5$,
for the $ (\Omega_0, \lambda_0)= (0.3,0.7)$
simulation, for $\Nmin=30$ by default, and for other values
of $\Nmin$ as indidcated.
The means are weighted by $\sin\theta$ 
and angles $\theta<5\deg$ are ignored.
Units are degrees. 
\label{t-angsp}}
$$\begin{array}{l ccc c ccc} \hline 
& \multicolumn{2}{c}{\Nmin=30}  & \Nmin=10 &\Nmin=30 &\Nmin=100 \\
z_i  & \mbox{H}& \mbox{D} & 
\multicolumn{3}{c}{\delta\theta \pm \Delta(\delta\theta)}
 \\ \hline
    6.0 &   25.08 &   24.25 &     0.8 \pm     1.6 
&    -0.8 \pm     1.8 &    -7.9 \pm     1.6\\
    5.5 &   25.02 &   15.31 &    -3.3 \pm     1.1 
&    -9.7 \pm     1.2 &   -13.5 \pm     1.4\\
    5.0 &   25.68 &   19.87 &    -0.4 \pm     0.8 
&    -5.8 \pm     0.9 &   -11.2 \pm     1.0\\
    4.5 &   25.99 &   22.75 &    -0.1 \pm     0.4 
&    -3.2 \pm     0.5 &    -5.5 \pm     0.6\\
    4.0 &   31.14 &   23.13 &    -4.5 \pm     0.3 
&    -8.0 \pm     0.3 &   -11.0 \pm     0.3\\
    3.5 &   33.79 &   38.97 &     6.5 \pm     0.2 
&     5.2 \pm     0.3 &     7.3 \pm     0.3\\
    3.0 &   34.20 &   35.62 &     3.7 \pm     0.2 
&     1.4 \pm     0.2 &     0.1 \pm     0.2\\
    2.5 &   39.05 &   43.22 &     5.1 \pm     0.1 
&     4.2 \pm     0.2 &     4.1 \pm     0.2\\
    2.0 &   37.43 &   45.18 &     7.5 \pm     0.1 
&     7.8 \pm     0.1 &     9.5 \pm     0.2\\
    1.5 &   38.12 &   45.20 &     7.2 \pm     0.1 
&     7.1 \pm     0.1 &     7.3 \pm     0.1\\
    1.0 &   36.76 &   42.42 &     5.9 \pm     0.1 
&     5.7 \pm     0.1 &     5.6 \pm     0.1 \\
    0.5 &   36.97 &   42.22 &     5.3 \pm     0.1 
&     5.3 \pm     0.1 &     5.1 \pm     0.1\\
\hline
\end{array}$$
\end{table}
}  

\def\tangve{
\begin{table}
\caption{Mean velocity 
angles $\theta_{\mbox{v}}$ (as in Fig.~\protect\ref{f-angve})
for non-debris halo (H) and debris (D) particles, their 
difference $\delta\theta_{\mbox{v}}$, 
and the $1\sigma$ uncertainty in the difference 
$\Delta(\delta \theta_{\mbox{v}})$  
[the uncertainty 
$\Delta(\delta\theta_{\mbox{v}})$ is 
estimated by using the standard
deviations of $\theta$, taking standard errors in the
mean, and assuming Gaussian error propagation],
as a function of redshift $z_i$, 
between $N$-body simulation output times at $z_i$ and $z_i -0.5$,
for the $ (\Omega_0, \lambda_0)= (0.3,0.7)$
simulation, for $\Nmin=30$. The means are weighted by $\sin\theta$ 
and angles $\theta<5\deg$ are ignored.
 Units are degrees.
\label{t-angve}}
$$\begin{array}{l ccc c c} \hline 
z_i  & \mbox{H}& \mbox{D} 
& \delta\theta_{\mbox{v}} \pm \Delta(\delta\theta_{\mbox{v}}) 
 \\ \hline
    6.0 &   50.13 &   52.42 &     2.3 \pm     2.4 \\
    5.5 &   50.87 &   47.40 &    -3.5 \pm     2.2 \\
    5.0 &   51.50 &   46.08 &    -5.4 \pm     1.0 \\
    4.5 &   50.28 &   48.36 &    -1.9 \pm     0.5 \\
    4.0 &   47.66 &   49.59 &     1.9 \pm     0.3 \\
    3.5 &   49.61 &   49.59 &     0.0 \pm     0.3 \\
    3.0 &   47.92 &   46.98 &    -0.9 \pm     0.2 \\
    2.5 &   46.79 &   47.39 &     0.6 \pm     0.2 \\
    2.0 &   47.67 &   47.17 &    -0.5 \pm     0.1 \\
    1.5 &   44.23 &   45.40 &     1.2 \pm     0.1 \\
    1.0 &   44.24 &   44.62 &     0.4 \pm     0.1 \\
    0.5 &   45.54 &   45.73 &     0.2 \pm     0.1 \\
\hline
\end{array}$$
\end{table}
}  

\def\trevir{
\begin{table}
\caption{Differing fates at an output time $t_{i+1}$ 
of matter which is `virialised' at output time $t_i$: 
inclusion in a primary successor or newly 
collapsed halo (H), non-virialised debris (NV), 
or re-virialised (RV, secondary successor halo)
debris (any particle can only be a member of one category),
 and the fraction $f \equiv N($RV$)/[ N($NV$) + N($RV$)]$
of debris which is `re-virialised', 
as a function of redshift $z_i$, 
between $N$-body simulation output times at $z_i$ and $z_i -0.5$,
for the $ (\Omega_0, \lambda_0)= (0.3,0.7)$
simulation. The total number of particles is $N=2097152$ and the
largest successor criterion is used as the merger/identity criterion
in order to maximise tracing of debris.
\label{t-revir}}
$$\begin{array}{l ccc c c} \hline 
z_i 
 & \mbox{H} & \mbox{NV} & \mbox{RV} & f 
 \\ \hline
6 &   7135 & 93 & 0 &           0 \\
5.5  & 14648 & 100 & 40 &       0.4 \\
5  & 28032 & 383 & 110 &        0.3 \\
4.5  & 47191 & 1444 & 755 &     0.52 \\
4  & 76986 & 4032 & 1785 &      0.44 \\
3.5  & 120674 & 8562 & 2006 &   0.23 \\
3  & 178627 & 15102 & 2346 &    0.16 \\
2.5  & 252710 & 23503 & 2456 &  0.10 \\
2  & 354118 & 30839 & 2651 &    0.09 \\
1.5  & 478918 & 46726 & 4440 &  0.10 \\
1  & 632984 & 61960 & 6045 &    0.10 \\
0.5  & 833417 & 61038 & 5618 &  0.09 \\
\hline
\end{array}$$
\end{table}
}  

\def\tmadau{
\begin{table}
\caption{Effect of merger/identity criterion choice on 
global star formation rates (cf Fig.~\protect\ref{f-madau}).
The decrease in star formation rate 
due to the exclusion of debris, 
as represented by using the 
{\em median} rather than the
{\em largest successor} merger/identity criterion,
is $\Delta_\psi(z) \equiv \log_{10}\psi(z,\mbox{\rm largest successor})$ 
$\;-\; \log_{10} \psi(z,\mbox{\rm median})$. 
See \SS\protect\ref{s-model} for the definitions of 
$\alpha$ and $\varepsilon,$ the efficiency parameters for 
the star formation
rate and reheating respectively.
\label{t-madau}}
$$\begin{array}{c c cc c c c} \hline 
\Omega_0 & \lambda_0 & 
\alpha & \varepsilon &
\Delta_\psi(z=1) & \Delta_\psi(z=3) 
& \max(\Delta_\psi) \\ \hline
%
%
\multicolumn{7}{l}{\mbox{\rm vary metric:}}\\
1.0 & 0.0 & 1.0 & 0.1 & 0.11 & 0.10 & 0.14 \\
0.3 & 0.7 & 1.0 & 0.1 & 0.04 & 0.11 & 0.12 \\
0.3 & 0.0 & 1.0 & 0.1 & 0.05 & 0.11 & 0.12 \\
\multicolumn{7}{l}{\mbox{\rm vary star formation rate:}}\\
0.3 & 0.7 & 0.1 & 0.1  & 0.11 & 0.12 & 0.14 \\
0.3 & 0.7 & 10.0 & 0.1 & 0.01 & 0.06 & 0.13 \\
\multicolumn{7}{l}{\mbox{\rm vary gas reheating:}}\\
0.3 & 0.7 & 1.0 & 0.01 & 0.07 & 0.12 & 0.13 \\
0.3 & 0.7 & 1.0 & 1.0 & 0.00 & 0.06 & 0.09 \\
\hline
\end{array}$$
\end{table}
}  

\def\tdlum{
\begin{table}
\caption{Effect of merger/identity criterion choice on differential 
luminosity 
functions $\phi(B)$ 
(shown in Fig.~\protect\ref{f-lfun} for $\Omega_0=1, 
\lambda_0=0$). The decrease in galaxy number density at characteristic
absolute magnitudes (`shoulder' and faint end of the luminosity function)
due to the exclusion of debris,
as represented by using the 
{\em median} rather than the
{\em largest successor} merger/identity criterion,
is $\Delta_\phi(B) \equiv \log_{10}\phi(B,\mbox{\rm largest successor})$ 
$\;-\; \log_{10} \phi(B,\mbox{\rm median})$.
\label{t-dlum}}
$$\begin{array}{c c c c } \hline 
\Omega_0 & \lambda_0 & \Delta_\phi (B=-21) & \Delta_\phi(B=-18) \\ \hline
1.0 & 0.0 & -0.02 & 0.22 \\
0.3 & 0.7 & 0.03 & 0.16 \\
0.3 & 0.0 & 0.03 & 0.16 \\
\hline
\end{array}$$
\end{table}
}  


\section{Introduction}  \label{s-intro}
An updated version of the galaxy formation modelling 
technique of \citet{RPQR97} is used here 
to examine one of the consequences of the fact that 
`(bottom-up) hierarchical'
galaxy formation is not strictly bottom-up hierarchical.

A bottom-up hierarchical galaxy formation scenario is 
(bottom-up) hierarchical in the sense that
small fluctuations in an otherwise constant matter density fluid 
initially collapse on small length and mass scales, forming
low mass dark matter 
haloes, which then merge together to form successively higher and 
higher mass
haloes\footnote{The word `halo' in this context is independent of
the word `halo' sometimes used to describe 
a spherically symmetric {\em stellar} component of 
the Galaxy, or of other galaxies.}. However, when haloes merge, 
significant fractions of the matter in pre-merger haloes 
are thrown out in tidal tails, bridges and other forms of debris. 
This has been well known since the 
review and further development of gravity-only $N$-body simulations of
close interactions of gravitating disks by \citet{TT72}.
Indeed, observations suggest that
at least some of these debris may form 
visible galaxies which are dynamically
bound entities \citep{DM94,Duc97}. Thus, the bottom-up hierarchy is
broken.

For recent discussions of tidal destruction of haloes in $N$-body simulations
see, e.g. \citet{Klyp99} or fig.~7 and fig.~8 of \citet{KK99}. For a debate
on the conditions required for the formation of tidal tail galaxies, 
see \citet*{Dub96}, \citet*{Mih98} and \citet{SpWh99}.

Ejection of matter into tidal debris is a top-down process.
So, bottom-up hierarchical halo formation is not 
strictly hierarchical: some material is (at least temporarily) 
thrown out of (or `down') the bottom-up hierarchy.

Bottom-up hierarchical\footnote{Hereafter, 
`hierarchical' means `bottom-up hierarchical' unless
otherwise specified.}
galaxy formation models have been 
modelled by several different techniques 
\citepf{RQP93,KWG93}. 

The primary output of hierarchical galaxy formation
models that can be compared with observations is 
the modelling of the stellar components of the 
haloes, i.e. galaxies. Stellar population models in hierarchical 
models depend on
star formation, or more precisely on star formation rate history
trees. In hierarchical galaxy formation models, 
star formation rate histories are physically best represented as trees,
since any single galaxy is (in general) the result of a history tree of 
galaxy merging 
\citep{RQP93,LC93,KWG93}. 

By assuming that haloes collapse in 
an idealised, monotonic, spherically symmetric way, 
 and that merging of haloes
proceeds monotonically in the sense that once matter is
considered collapsed (virialised) in a halo, it remains
collapsed when the halo merges with other haloes,
\cite{KWG93} used the probabilistic formulae 
of the Press-Schechter formalism (e.g. \citealt{PS74,LC93})
to generate merging history trees of haloes, and followed through to
merging trees of galaxies and to synthetic values of galaxy 
observables such as apparent magnitudes.

\cite{RQP93} generated merging history trees without making this
physical assumption by directly detecting haloes in 
pure gravity $N$-body simulations, and then followed through to
galaxy merging trees and apparent magnitudes.

What differences can arise by avoiding the monotonic collapse
assumption of the Press-Schechter formalism?
The star formation rates in the material
ejected as tidal debris are unlikely to be the same as 
those which would be expected if the material fell in smoothly 
according to the monotonicity assumption.

Could the monotonic collapse assumption be a good approximation
if
\begin{list}{(\roman{enumi})}{\usecounter{enumi}}
\item the ejected material remains gravitationally 
bound to the merging system, i.e., 
it forms a low-density `atmosphere' 
around the halo resulting from the merger (the merger product), 
and/or 
\item it falls back within a dynamical time?
\end{list}

In both these cases, 
the physical conditions
of the cold gas fraction of these debris
are likely to be temporarily different from those of the
 gas which 
falls directly into a disk at the centre of the main merger product
without passing through this `debris' phase. Hence, the star
formation rate, which is observed to depend on density 
(e.g. \citealt{Kenn98}), is likely to be different 
for matter which passes through a `debris' stage compared to
matter falling smoothly into a disk. 

The difference in the star formation rate of debris 
relative to that for the monotonic infall case
could, in principle, be either negative or positive.
Gas in low-density streams or atmospheres is likely to form stars
more slowly than in the disk, whereas gas inside newly 
condensed tidal dwarf haloes may be of higher density than
in the disk and form stars much more rapidly.

However, 
stars formed in the debris are likely either to dynamically
become part of the main merger product, or to observationally 
be counted in the integrated luminosities and colours of the
main products in faint galaxy catalogues. It may then be 
a valid approximation (for some galaxy statistics) 
to ignore the violation of hierarchy, 
i.e. to assume that the debris provide dark matter (non-baryonic 
plus baryonic) which is available for gas cooling and star formation
as if they were part of the virialised successor halo, 
as if they were {\em not} thrown out of the 
successor halo.

As mentioned above, this is the assumption implicitly made in 
semi-analytical galaxy formation models based either on Press-Schechter
probability estimates of hierarchical halo formation 
(e.g. \citealt{KWG93,GHBM98}) or on extrapolation from linear
density fields (the block model of \citealt{ColeGF94}; 
the multi-cell-merging model of \citealt{RT96}, see 
also \citealt*{Naga98,LMG00}). 
Following the implicit assumption of monotonic, spherical infall, 
assumptions on gas cooling, star formation rates and  
gas reheating are made, in order to derive observational
properties of galaxies.

The two limiting cases of the fate of debris which should bound 
other possibilities
can be stated in simple terms as:
\begin{list}{(\alph{enumi})}{\usecounter{enumi}}
\item most debris either becomes dynamically independent or
is `effectively independent' in that 
it takes a large fraction of a Hubble time to fall back into the
virialised part of the halo product and does not contribute stars to
the central galaxy from an observer's point of view;  or
\item most debris either forms an `atmosphere' close to the central
galaxy or falls back into the central galaxy relatively 
quickly so that the mononotonic,    
spherical infall assumption is a good approximation
and star formation happens as if infall really were monotonic.
\end{list}

How much effect could the 
difference between (a) and (b) have on 
galaxy statistics such as the luminosity function?
What would be the decrease in star formation rates and how
much could this affect the present-day luminosity function?

These questions can be investigated by looking at the effect
that the merger/identity criterion used in merging history tree
galaxy simulations can have on the luminosity function.

This is the purpose of this paper. 
An $N$-body plus semi-analytical galaxy formation modelling
technique \citep{RQP93,RPQR97} is used to study the effects
that tidal debris could have on global galaxy statistics.
The technique used is that of combining 
gravity-only $N$-body simulations with semi-analytical formulae
\citep{RQP93,RPQR97}, 
updated from the \citet{RPQR97} version by the
creation of a multiple-galaxy-per-halo merging history tree using the
dynamical friction formalism of \citet{LC93},
and the inclusion of assumptions 
on gas cooling, star formation rates and  
gas reheating as in \citet{KWG93}.
A brief description of the updated model, which is publicly available as 
the {\arfus} software package, is provided in the Appendix.

Use of this galaxy formation model enables investigation of 
the negative effects that the passage of matter through a
`debris' stage could have on star formation rates and luminosities.

In \citet{RPQR97}, the problems of tidal debris were noted and a
simple, robust `merger/identity' 
criterion for deciding how to identify halo progenitors
and successors between successive output time steps was adopted. 
In \SS\ref{s-crit}, ways to modify this merger/identity criterion 
and the way in which these modifications enable investigation of
the limiting cases (a) and (b) for the effects of debris on 
galaxy formation are presented.
In \SS\ref{s-effets}, effects on galaxy statistics
are shown. 
The results are discussed and conclusions provided 
in \SS\ref{s-concl}.

Only one cosmological model (the perturbed Friedmann-Lema\^{\i}tre
model) is considered here.  The Hubble constant is parametrised here as
$h\equiv H_0/100$\,km~s$^{-1}$~Mpc$^{-1}.$ Three options for the other
metric parameters (the present values of the 
density parameter, $\Omega_0$, and of the dimensionless cosmological
constant, $\lambda_0$) are considered (see Table~\ref{t-median}).

\section{The merger/identity criterion and debris} \label{s-crit}

The principle of combining 
$N$-body simulations with semi-analytical techniques, in which 
a series of output times from the simulations are used, 
was presented in \citet{RQP93,RPQR97} and has more recently been adopted
by \citet{KCDW99a}. 

This should not be confused with the technique
of applying the extended Press-Schechter formalism to create merging histories
for haloes detected in {\em one} $N$-body simulation output time 
\citepf{KNS97,Gov98,Bens99}. The latter could possibly be described as using
$N$-body simulations to `normalise' the Press-Schechter technique of
generating a set of merging history trees, as opposed to using the
$N$-body simulations to generate the trees directly. 

The precise algorithm for constructing the halo merging history 
tree is described in \SS2.1.3 of \citet{RPQR97}. 
In that paper, 
haloes between two 
successive $N$-body simulation output times $(t_i,t_{i+1})$ 
were considered to have a merger/identity link if
at least $P=50\%$ of the particles in 
a halo at $t_i$ were present in a halo at $t_{i+1}. $ 
This is the {\em $P=50\% $ criterion}.
With this criterion,
and for the same halo detection method (group finder algorithm) 
at each output time, some haloes
were found not to have successors. That is, 
there are some haloes for which more than 50\% of their
material forms tails, cusps, shells and low density
envelopes outside of the group-finder defined boundaries of haloes.

Given this criterion, typically $P \sim 75-90\%$
of a halo at $t_i$ is present in a halo at $t_{i+1}$ (table~3, 
\citeauthor{RPQR97}). This fraction depends on the group finder
density detection threshold and on the slope of the power spectrum
of density perturbations used in the $N$-body simulation.

\citet{KCDW99a} adopted a slightly modified version of this $P=50\%$ criterion 
as their merger/identity criterion, but haloes failing the criterion 
were considered to generate `lost' galaxies which could later on 
either merge with a central galaxy or become a satellite galaxy if the
criterion were satisfied at a later $N$-body output time.


Both the original criterion of \citeauthor{RPQR97} and that of 
\citeauthor{KCDW99a} are bracketted by the following two limiting
cases.

\subsection{The {\em `median'} merger/identity criterion for modelling
case (a): the loss of debris} \label{s-median}

By terminating and (in effect) removing haloes
that fail the $P=50\%$ criterion from later times in the halo merging 
history tree, the ejection of debris is implicitly modelled.
However, the effect is in fact exaggerated, since
stars previously formed in haloes that fail the
$P=50\%$ criterion 
no longer contribute to the stellar population of any galaxy, neither
in the present timestep (as they are no longer in haloes) nor later
(as they are not added to future haloes).
This is useful for obtaining an upper bound to the
loss in star formation and in luminosity that would be caused by the
debris stage, i.e. to bound (a) from above.

However, the choice of 
$P=50\%$ as the minimum merger/identity percentage
is an arbitrary (though simple) choice designed to guarantee at most
a single successor halo, which does not have any other physical motivation.
A less arbitrary choice for a merger/identity 
criterion should be one
that both
\begin{list}{(\roman{enumi})}{\usecounter{enumi}} 
\item excludes haloes of which the fraction thrown into 
debris is higher than `typical', and 
\item retains the property of guaranteeing a single successor halo.
\end{list}

The choice of the successor halo that contains 
the highest fraction of a progenitor
halo is sufficient to satisfy (ii), except for the physically unlikely 
but numerically possible case where a progenitor halo contributes
two equal (rational) fractions (e.g. 11/30ths) to two successor 
haloes.

For property (i),
the median merging fraction ($\pmed$) can be used to define
a `typical' merging fraction, so that mergers for which the
debris fraction is `much higher' than the median debris 
fraction, i.e. $1-P \gg 1-\pmed$, can be excluded. 

An obvious way to define `much higher' is to note
that all mergers with percentages {\em higher} than the median
($P > \pmed$) 
ought to be included by the criterion. By choosing an exclusion percentage
which is as far below the median as 100\% is higher than the median,
i.e. for which 
\begin{equation}
1-\pmed = \pmed - P
\;\;\;\Leftrightarrow \;\;\;
P=2\pmed -1,
\label{e-ppmed}
\end{equation}
a symmetry about the (original) median is obtained.

This defines the {\em `median' merger/identity criterion}:
a halo at $t_i$ is
linked with the halo at $t_{i+1}$ that contains 
the largest fraction of particles from the halo at $t_i,$ 
as long as that successor halo (at $t_{i+1}$) contains more than
a fraction $P=(2{\pmed} - 1)$ of the particles in the halo at $t_i,$
irrespective of whether the merging fraction
is greater than 50\% or not. 

The median fraction $\pmed$ is calculated
from the full set of output time pairs for a given $N$-body 
simulation, and for reasons of dynamical consistency is likely to remain well
above 50\%. The value $P=(2{\pmed} - 1)$ may, in principle,
fall below 50\%. If it is above 50\%, then the rare cases of equal 
contributions to multiple successor haloes mentioned above are automatically
excluded.

The algorithm in \SS2.1.3 of \citet{RPQR97}
can easily be modified in order to implement this (and the
following) merger/identity
criterion, by sorting the two arrays
of halo numbers $a_i,$ $a_{i+1}$ in such a way that the array
$a_i$ (instead of the array $a_{i+1}$) becomes a non-decreasing
arithmetical sequence.

\subsection{The {\em `largest successor'} criterion for modelling
case (b): smooth, monotonic infall} \label{s-largest}

\tmidcrit

For the other limiting case, in which debris is to be included 
as much as possible in successor galaxies, 
the {\em `largest successor'} 
merger/identity criterion is defined as follows.

Each halo at $t_i$ has
a link made with the halo at $t_{i+1}$ that contains the largest 
fraction of the particles in the halo at $t_i,$ even if this is fraction
is less than 50\%. This requires that at least $N=1$ particle in a halo
at $t_i$ is present in a halo at $t_{i+1}$ in order for the halo at
$t_i$ to have a successor. This is usually the case. 

As in the case of the $P=50\%$ and median criteria, a halo has
at most one successor with the largest successor criterion, 
since, by definition,
only one halo at $t_{i+1}$ can have the {\em largest} fraction 
of particles of the halo at $t_i$, apart from the unlikely
case of equal rational fractions, as mentioned above. In the $N$-body
simulations studied here, the latter case (double largest successor, 
with a merging percentage of less than $\sim 20\%$),
occurs in about 0.1\% 
of cases, and an arbitrary choice between the two successors is made.

The different merging/identity criteria are summarised in 
Table~\ref{t-midcrit}.

\subsection{$N$-body simulations and group-finding} \label{s-nbody}

The $N$-body simulations used are
particle-mesh $N=2\e{6}$ 
simulations in a comoving `periodic box' of size 32{\hMpc}
[i.e. the cosmology is hypertoroidal, e.g. \citet{FMel,LR99}]. 
Three choices of metric parameters are used:
$(\Omega_0=1,\lambda_0=0,h=0.65),$ 
$(\Omega_0=0.3,\lambda_0=0.7,h=0.65)$ and
$(\Omega_0=0.3,\lambda_0=0.0,h=0.65).$ 
The baryon density is $\Omega_b=0.05.$
The initial perturbation spectrum is a CDM-like 
spectrum [eq.(2) of \protect\citet{DEFW85}, where 
$\Gamma\equiv \Omega_0 h $ is substituted for $\Omega_0 h$], 
where $\Gamma=0.3$ and the normalisation at $8${\hMpc} is
$\sigma_8=0.6$.

Haloes are detected using the University of 
Washington friends-of-friends (FOF) 
group-finder\footnote{Available at 
{\em http://www-hpcc.astro.washington.edu/tools/}.}
 with a friendship distance of $b=0.2$
times the mean interparticle spacing and a minimum of 30 particles
per halo.

\subsection{Star formation assumptions for comparing the median and
largest successor merging/identity criteria}
\label{s-model}

To see the way that the 
median (\SS\ref{s-median}) and 
largest successor (\SS\ref{s-largest}) merging/identity criteria
may respectively 
test the limiting cases (a) and (b) (\SS\ref{s-intro}), the
halo merging history trees resulting from the different criteria
can be combined with a semi-analytical formalism to derive
star formation and luminosity statistics. This is done here using
the galaxy formation model
{\arfus}, updated from \citet{RPQR97}, which applies 
semi-analytical star formation hypotheses to a multiple-galaxy-per-halo
merging history tree. {\arfus}-V0.03 is used in the present study, 
with the simple stellar populations (for solar metallicity and a 
Salpeter initial mass function) of \citet*{DGS99}.

The model presented in \citet{RPQR97} represented maximal galaxy
merging, i.e. there was a 1:1 correspondence between 
haloes and galaxies and the galaxy merging history tree was identical
to the halo merging history tree. In the updated model, 
a galaxy merging tree is derived from the halo merging tree
by following \citeauthor{LC93}'s (1993) suggestion:
\begin{list}{(\roman{enumi})}{\usecounter{enumi}}
\item when two haloes merge, the galaxy in the more massive halo
becomes the central galaxy of the successor halo, while the galaxy
in the less massive halo becomes a `satellite' galaxy, and
\item the satellite galaxy is allowed to spiral in 
to the centre of the parent halo by dynamical friction.
\end{list}
The dynamical friction merger time is estimated via 
eq.~(4.2) of \citet{LC93}, including the effect of orbital eccentricity.

Note that the merger time of the satellite galaxy may be 
greater than the interval between $N$-body simulation output
times. This leads to an increased complexity in both calculation
and storage of the combined galaxy plus halo merging history tree.
See the Appendix (Sect.~\ref{s-afhtog}) for the algorithm adopted
in {\arfus}-V0.03.

As in \citet{KWG93}, cooling functions 
and singular isothermal total matter
density profiles are used to calculate cooling radii within which gas
cools, a star formation rate depending on the amount of cold gas and
the dynamical time is applied, and some gas is reheated by energy from
supernovae. The different options (a) and (b) for the halo merger/identity
criterion lead automatically to different amounts of gas available for
star formation, and to loss of stars from debris.

The cooling and heating rate formulae of \citet{KWG93} are adopted here,
with a standard value of 
the reheating efficiency free parameter, $\varepsilon =10\%$ 
[eq.~(9) of \citet{KWG93}] and 
the cooling functions of \citet{SuthD93}. The star formation rate
is 
\begin{equation}
{dM_* \over dt} = \alpha 
{\Mcold  \over \tdyn},
\label{e-sfr}
\end{equation}
where $M_*$ is the stellar mass, 
$\alpha$ is a star formation rate efficiency free parameter normally
set to $\alpha=1$, 
$\Mcold$ is the mass of cold gas and
$\tdyn$ is the dynamical time of the parent halo at the earlier 
$N$-body simulation output time of the output time pair.

See the Appendix for more comments on {\arfus}, or the
electronic site of {\arfus} (address in Appendix) 
for full documentation.

\section{Results} \label{s-effets}

\subsection{Merging fractions}

\ffrac
\fpmeddelz
\fmedian

The relevance of the merging/identity criteria
can be verified by studying the following questions.
\begin{list}{(\roman{enumi})}{\usecounter{enumi}}
\item How does the fraction of debris depend on halo mass?
\item How do they depend on the output time (redshift) and the
output time (redshift) intervals? 
\item How far from the central halo are debris located?
\item Do the debris just form loose atmospheres and tidal tails,
or do they revirialise and form tidal dwarf haloes/galaxies?
\end{list}

\tmedian

\tmedianPoiss


Figs~\ref{f-frac}-\ref{f-rmedian} and Table~\ref{t-median} provide
responses to these questions for some typical $N$-body simulations. 

\subsubsection{(i) How does the fraction of debris depend on halo mass?}
Fig.~\ref{f-frac} clearly shows that for a flat, critical density
CDM-like universe, 
the low mass haloes are more disrupted than the high mass haloes.
This is physically reasonable, since it is more likely that 
low mass haloes are 
partially broken up by high mass haloes than the inverse. Similar
results are found in the $N$-body simulations with different 
values of the 
metric parameters $\Omega_0$ and $\lambda_0$ 
(and also in simulations with different initial
perturbation spectra).
In the present case, a conservative
minimum of $\Nmin=30$ particles per halo was used in application of the
friends-of-friends (FOF) group finding algorithm, so this is unlikely to
be an effect of resolution.

Nevertheless, the merging percentages of many haloes more massive than
$10^{12}M_{\sun}$ are well below $2\pmed-1,$ where 
$\pmed \approx 90\%$ from Table~\ref{t-median}, i.e. well below $80\%$.
This confirms 
the utility of studying the possible effects of the
break-up of these haloes on star formation.

Could it be useful to consider analyses with smaller or
larger minimum numbers of particles per halo? While the
calculation 
of Poisson errors of the statistics of interest here
would not be simple, these can nevertheless be discussed at an order
of magnitude level.

The Poisson 
uncertainty in a halo of $\Nmin=30$ particles is $\sqrt{30}$, i.e.
a fractional error of $18\%$. This is of similar order of magnitude to
the median losses represented in Fig.~\ref{f-frac} and 
Table~\ref{t-median}. 

However, most haloes have masses considerably greater than the
minimum required, so while uncertainties due to Poisson noise 
at the low mass cutoff may be important for the 
smallest haloes, it is unlikely to be globally important.
So, increasing the cutoff criterion to $\Nmin=100$ particles should
not greatly affect the analysis.
Table~\ref{t-medianPoiss} confirms that median merging fractions
are affected only slightly. 

Table~\ref{t-angsp} (Sect.~\ref{s-rmedian}) shows that more subtle 
statistics, such as relative orientations of debris and non-debris
particles' vectors to successor halo centres, are modified 
somewhat more, particularly at earlier redshifts. This is to be
expected, since at early redshifts, high mass haloes do not yet
exist, so more haloes are closer to the cutoff limit at high redshift
than at low redshift.

On the other hand, decreasing the cutoff, to $\Nmin=10$ 
particles, for example,
is certainly going to increase Poisson noise: $33\%$ is a high 
uncertainty even for estimates of first order statistics.

Table~\ref{t-medianPoiss} supports this argument, 
showing that lowering from $\Nmin=30$
to $\Nmin=10$ has considerably greater effect (at $z \le 3$)
than that of increasing to $\Nmin=100$.

Therefore, the criterion of $\Nmin=30$ appears to be within the region
of convergence of the most basic debris statistic as a function of
$\Nmin$. The more subtle statistics shown in Table~\ref{t-angsp}
support $\Nmin=30$ as being closer to a domain of convergence than
$\Nmin=10$.

Hence, results below are all for $\Nmin=30$ unless otherwise
specified.

\subsubsection{(ii) How do the debris 
fractions depend on the output time (redshift) and the
output time (redshift) intervals?}
Table~\ref{t-median} shows that, in general,
the median merging percentages
start close to 100\% and decrease as time increases. It was remarked
upon above that the break-up of low mass haloes by high mass haloes
is more physically reasonable than the opposite. If this is a necessary
condition for the break-up of haloes, then it is necessary to have
a large dynamic range in halo masses in order for the mass ratios to be
large enough for the effect to occur. 

For a given $N$-body simulation 
in which halo formation is (bottom-up) hierarchical, and which
has a fixed mass resolution independently of time, the first haloes
to collapse necessarily have roughly equal masses, just slightly greater
than the detection threshold. It is only at later times that high mass
and low mass haloes can coexist, and so, as is suggested here, 
the high mass haloes can tend to cause the low mass haloes to `lose' 
some of their mass outside of halo boundaries defined by 
a given group-finding criterion.

This provides one possible explanation for why 
the median merging fractions can decrease with time as shown
in Table~\ref{t-median}.

An alternative explanation can be expressed in terms of dynamical
times. The fraction of debris loss $1-P$ should be related in 
some way to the fraction of a dynamical time represented by an
interval $\Delta z$. Since a fixed (virialisation) overdensity threshold
is used to detect haloes, 
\begin{equation}
\tdyn \propto \rho_{\mbox{\rm \small halo}}^{-1/2} 
\propto \left<\rho\right>^{-1/2} 
\propto (1+z)^{-3/2},
\end{equation}
so that 
\begin{equation}
{ {\rm d}\tdyn \over \tdyn } 
\propto 
{ {\rm d}z \over (1+z) }
\end{equation}
Hence, a fixed choice of interval $\Delta z$ corresponds to 
fractions of a dynamical time
$ {\rm d}\tdyn / \tdyn $ which increase with increasing time 
(decreasing $z$). 

The interpretation of the time dependence of $\pmed$ 
in Table~\ref{t-median} is therefore that at early times, $\Delta z$
corresponds to dynamical time fractions that are too short to have
time for much debris loss. As $ {\rm d}\tdyn / \tdyn $ increases,
the loss fractions increase ($\pmed$ decreases), 
until  $ {\rm d}\tdyn / \tdyn $ becomes
large enough that matter has time to fall back into the haloes, in
which case the loss fraction decreases ($\pmed$ increases).

The latter effect is also shown by the dependence on redshift {\em
interval} revealed in Table~\ref{t-median}.  This shows that
increasing the time interval (at constant redshift) enables the debris
from progenitor haloes to mostly fall into the successor haloes. This
implies that {\em although the problem of debris may temporarily modify the
monotonic, spherically symmetric infall assumption, the effect is
limited in time}.

Table~\ref{t-median} also shows that 
increasing the time interval (redshift interval) shifts the 
minimum in $\pmed$ to earlier times. This is consistent with the
above explanation.

\fangsp
\fangve

\tangsp
\tangve

If the redshift interval is made {\em shorter} than the values shown in 
Table~\ref{t-median}, then the time resolution becomes smaller than
typical dynamical times, so that debris does not have time to be lost
between successive $N$-body output times. Between the two extremes of
small intervals over which debris does not have time to be ejected, 
and large intervals over which there is sufficient time for debris to
fall back into the successor haloes, a minimum in $\pmed$ 
(a maximum debris loss) must occur. 

Fig.~\ref{f-pmeddelz} shows this for the $(\Omega_0=0.3, 
\lambda_0=0.7)$ case. The maximum debris loss occurs at 
$\Delta z \approx 0.4$, and the two dependences of debris loss 
above and below the minimum can be estimated as:
\begin{equation}
\pmed \approx \left\{ 
  \begin{array}{lll}
  81\% -9\% \log_{10}(\Delta z), & 0.025 \ltapprox \Delta z \ltapprox 0.4 \\
  94\% +21\% \log_{10}(\Delta z), & 0.4 \ltapprox \Delta z \ltapprox 2 
  \end{array}
  \right. ,
\label{e-pmedian}
\end{equation}
though clearly these relations can only hold over approximately 
the intervals
stated, since for small intervals, 
$\pmed \rightarrow 1^{-}$ as $\Delta z \rightarrow 0^{+}$ and for large 
intervals, $\pmed \le 1$ implies that the relation cannot continue
above $\pmed =1$.

It should also be noted that the formal uncertainties on $\pmed$
diminish as $\pmed \rightarrow 1$, since if the number of particles
in a progenitor halo is $N$, then the binomial uncertainty
in $N\pmed$ is $\sqrt{N\pmed(1-\pmed)}$, so that
\begin{eqnarray}
\Delta\pmed    &=& \Delta [(N\pmed)/N] \nonumber \\
              &=& \Delta [N\pmed]/N \nonumber \\
                 &=& \sqrt{N\pmed(1-\pmed)} \; / \; N \nonumber \\
                 &=& \sqrt{\pmed(1-\pmed)}  \;/ \; \sqrt{N} \nonumber \\
                 &\rightarrow& 0 \;\mbox{as}\; \pmed \rightarrow 1.
\label{e-binomerr}
\end{eqnarray}

Comparison with table~3 of \citet{RPQR97} (which shows {\em mean} values 
of $P$, for less conventional 
choices of group-finding detection thresholds) indicates that the 
various $P$ estimates will vary somewhat 
with choice of group-finder and initial perturbation spectrum.

However, presumably 
because eq.~(\ref{e-pmedian}) is based on a redshift interval rather
than a time interval, it appears nearly independent of the
metric parameters $\Omega_0$ and $\lambda_0.$

\fttail
\fttailB
\fttailC

\subsubsection{(iii) How far from the central halo are debris located?}
\label{s-rmedian}
Fig.~\ref{f-rmedian} shows that, in general, the debris is 
located close to the successor haloes. This is reassuring for those
who wish to ignore the debris. In fact, the median distances of 
particle debris are mostly {\em closer} to the centre of the 
successor halo than its outermost radius. This would not be possible
if a spherical overdensity group-finder were used, but since 
dark matter haloes are generally found to be triaxial [in the sense
that they are {\em not}, in general, spherically symmetric, 
e.g. \citet{Warr92,LC96}], this simply means that the vectors from 
debris to the successor halo centre are less aligned with 
the major axis of the successor halo than are the vectors from
non-debris halo particles. This suggests that the debris are
also more likely to be
in orbit around one (or both) of the minor axes of the successor
halo than non-debris particles.

These explanations are examined quantitatively at the epoch 
of largest debris loss in Figs~\ref{f-angsp} and \ref{f-angve},
and as a function of time in 
Tables~\ref{t-angsp} and \ref{t-angve}.

Overall, these figures and tables clearly show that the debris
particles' position vector distributions 
have significantly different spatial orientations to those of the
non-debris particles by several degrees, 
relative to the halo major axes. Given the formal errors 
listed, some of the velocity distributions would also appear to 
have significant differences, but the lack of a smooth trend
suggests that this may just be due to noise.

So, while the spatial alignment of debris relative to non-debris
is confirmed to avoid the major axes, as was suggested above 
to explain Fig.~\ref{f-rmedian}, kinematic differences in the
two populations appear to be very small.

Moreover, Table~\ref{t-angsp} quantitatively shows the shift
from mostly filamentary structures at high redshifts to much
more symmetric structures at later times. It also shows that
the debris orientations 
are {\em closer} to the major axes at earlier times.
This filamentarity 
starts from around $10\deg$ more `filamentarity' at $z_i= 5.5$
(ignoring $\theta$ 
for the $z\in [6.0,5.5]$ interval, which is insignificantly 
different from zero), and decreases and switches through to a maximum 
debris `oblateness' approximately coinciding with the epochs
of maximum debris loss at the same time as the non-debris material
becomes more spherically symmetric.

\trevir

Another property revealed by Fig.~\ref{f-rmedian} 
is that the general slope of the relation in the figure 
shows that the higher the fraction
of a progenitor halo is expelled as debris (i.e. the lower the 
merging fraction), the more likely this debris is to be further from 
the centre of the successor halo. This is dynamically unsurprising: the
less bound the progenitor is to the successor, the lower the merging 
fraction is likely to be and the further matter is likely to be ejected
from the successor.

While these subtleties of tidal debris dynamics are unlikely to have
significant effects on global star formation statistics at the broad
level studied in this initial exploration of tidal debris effects,
these results could presumably be taken into account in future
second order studies.

\subsubsection{(iv) 
Do the debris just form loose atmospheres and tidal tails,
or do they revirialise and form tidal dwarf haloes/galaxies?}
\label{s-ttail}

Figs~\ref{f-ttail}, \ref{f-ttailB} and \ref{f-ttailC} 
show an example of material classified as debris in one of 
the $N$-body simulations studied, using the largest 
successor criterion in order to maximise the number 
of particles accounted for as debris.

Quantitatively, material which is merely low density
is distinguished from material which has `re-virialised', i.e.
which is detected as part of a secondary halo at the later output time 
($z=1.0$ in the case illustrated). The former is traced by large
gray dots and the latter by large black dots. 

Qualitative distinctions in the different forms of debris can
be made by visual inspection of the figures.  The low density material
(large gray dots) appears to form a loose atmosphere, while the
`re-virialised' matter appears to form several temporary 
`shells' (bunches of
large black dots at the bottom left and at the top of the $X-Y$ view
of the main halo in Fig.~\ref{f-ttailB}, i.e. at $z=1.0$). 
Shells are known to form as 
a general consequence of merging bodies under gravitation, and have
been observed in optical images of galaxies
\citepf{Malin83,Q84,HQ88,HQ89}.

In the latter category, is it useful to try to distinguish genuinely
re-virialised matter, i.e. tidal dwarf haloes which are bound
objects, from the temporary phenomena of tidal tails and shells which
recollapse to the main halo?

The analyses of 
\citet{Dub96}, \citet{Mih98} and \citet{SpWh99} suggest that whether
or not tidal tail dwarf haloes/galaxies can be formed in models of
merging gravitational discs depends on the details of the galaxy 
formation recipe adopted. 

Since dissipation is not included in the
pure gravity, $N=2\e{6}$-body simulations studied here, realistic discs
are unlikely to form and robust constraints on virialised tidal dwarfs
cannot be expected.

Moreover, observations \citep{DM94,Duc97,Mend00} imply masses of tidal
tail galaxies of around $10^{9}-10^{10} M_{\sun}$, 
a scale which is well below that of 
masses corresponding to $L^*$, the
characteristic `shoulder' mass of the galaxy luminosity
function, of around $10^{11}-10^{12} M_{\sun}$. 
These tidal dwarf masses are also below the 30-particle
minimum mass of haloes detected in the present simulations: $6\e{10}
M_{\sun}$ (for the $\Omega_0=0.3$ simulations, where particle mass
is $2\e{9}M_{\sun}$).

In any case, the amount of `re-virialised' matter, whether in 
unbound tidal tails and shells or in bound tidal dwarfs, is small.
This result is shown in Table~\ref{t-revir}, for the 
$\Delta z=0.5$ interval, which gives close to maximum debris
fractions.

Comparison of Tables~\ref{t-median} and \ref{t-revir} shows 
that once the amount of debris becomes non-negligible, i.e. 
10\% or greater (during the $z\in [3.0,2.5]$ interval for the
non-zero cosmological constant metric), the fraction of this debris
which is comprised of re-virialised matter approaches around 10\%.
That is, the fraction of virialised matter at one time step which
does not contribute to the most massive successor halo of a given halo,
but instead to a secondary `halo', is only around $\sim 1\%$ or less.

This is consistent with the expectations from observational limits
on tidal dwarf formation, and suggests that tidal dwarfs are unlikely
to contribute significantly to first order global galaxy statistics,
even if all tidal tails and shells which passed through dense stages
were considered to form independent, bound systems.

\fmfun

\fmadau

\subsection{Mass functions}

Figure \ref{f-mfun} shows that use of the stricter
merger/identity criterion, i.e. the $P=78\%$ {\em median} criterion 
instead of the {\em largest successor} criterion, 
causes a strong effect in the low mass haloes and little (global) 
effect in the high mass haloes. 
This is consistent with what 
can be expected from Fig.~\ref{f-frac}, since the haloes which 
have less than $P=78\%$ merging fractions are mostly,
but not all, of small masses.

Note that these mass functions 
are constructed from output times independent of the original 
$N$-body simulation output times. Sets of haloes are detected in 
individual $N$-body output times, a halo merging history tree is 
then constructed, a galaxy+halo merging history tree
is calculated, and finally a star formation rate history tree 
is evaluated.
The mass functions shown represent the parent haloes of all galaxies
present in the star formation history tree at the chosen (arbitrary) 
output time which existed as individual haloes at the $N$-body 
simulation output time $t_i$ preceding the chosen output time.
This provides a double check on the interpolation
and merging time algorithms, since the set of haloes represented here 
in a mass function should be a subset of the haloes present at $t_i$
and a superset of the haloes present at $t_{i+1}.$ 

\tmadau

With the application of the median merger/identity criterion, 
it is often the case that
a halo at a time $t_i$ has no successor at time
$t_{i+1}$ because of the strictness of the criterion ($P > 2\pmed-1 $ 
for the median criterion compared to $P > 0$ for the largest
successor criterion).
In this case, the halo's mass does not contribute to any branch of 
the tree during the
interval $(t_i,t_{i+1}).$ It may have contributed to the tree during
$t<t_i,$ and some of its constituent total matter particles are likely to
contribute to haloes later, during $t>t_{i+1},$ but during 
the interval  $(t_i,t_{i+1})$ it is considered to be non-existent
as far as the star formation rate history tree is concerned.

The physical interpretation of this in terms of mass distribution
is that the material from a halo is being `smeared out' as the halo
passes through or is torn up by its companion halo(es). 

How wrong, in terms of implied star formation or luminosity estimates,
is the Press-Schechter approximation in its ignorance of this process?
The estimates made here are as follows.

\subsection{Star formation and luminosity functions}


\flfun

\tdlum

Figs~\ref{f-madau} and \ref{f-lfun} 
show how the effect of star formation truncation
in haloes which turn into `debris' translates to 
global star formation rates and luminosity functions.

As discussed in \SS\ref{s-nbody} and \SS\ref{s-model},
this is carried out using {\arfus}-V0.03,  
where adoption of the 
median (\SS\ref{s-median}) and 
largest successor (\SS\ref{s-largest}) merging/identity criteria
test the limiting cases (a) and (b)
respectively 
as defined in \SS\ref{s-intro}.

The principal effect in the luminosity functions 
is that the reduction in the number of low
mass haloes is visible as a reduction in the number of low luminosity
galaxies, and overall as a reduction in the global star formation 
rate. Table~\ref{t-dlum} shows the extent to which the reduction in 
the number of low luminosity galaxies depends on the metric parameters
$(\Omega_0, \lambda_0).$ There is some dependence on the matter density,
i.e. the reduction in number density is less strong for lower matter
density, but there is 
negligible dependence on the cosmological constant.

The main effect in the global star formation rate (Fig.~\ref{f-madau})
is a difference of 
about 0.05--0.10~dex (see Table~\ref{t-madau})
at intermediate redshifts ($1\ltapprox z \ltapprox 3$). 
Given the uncertainty in 
observational estimates at super-unity redshifts, this is presently
an effect smaller than the observational uncertainty. 

Table~\ref{t-madau} shows that this effect has little dependence
on either the metric parameters or on particular choices of the
efficiency parameters for the star formation rate and gas reheating.
Because of the discreteness in the redshifts of the $N$-body simulation
output times, the max$(\Delta_\phi)$ values are more robust estimates
of the effect of debris than the $\Delta_\phi(z=1)$ and 
$\Delta_\phi(z=3)$ values, particularly in the cases of $\alpha=10.0$
and $\varepsilon=1,$ in which excessively high star formation rates or
reheating (respectively) 
cause sharp oscillations in the global star formation rates related
to the $N$-body output times.

In can be seen in Fig.~\ref{f-madau} that 
at low redshifts, 
the observational uncertainties quoted are mostly smaller than about 
0.1~dex. If the
\citet{Lilly96} estimates are assumed to suffer a systematic error,
then the apparently smooth behaviour of the other estimates would 
imply that the {\em differences}
in the global star formation rates for the two 
merger/identity criteria are greater than the uncertainties
in the low redshift observational estimates.

However, since the two rates approach each other as $z\rightarrow 0$ 
for any of the choice of metric parameters, the model difference 
again becomes smaller than the observational uncertainty. 


Why should the global star formation
rates for the two merger/identity criteria approach each other towards
the present epoch?
Given that the median merging percentages do not suddenly approach 
unity in the lowest redshift intervals (though they do start to increase
slightly, see Table~\ref{t-median}), this cannot be because matter is
no longer thrown out into debris at late epochs.
A possible explanation is
the decrease in the contribution of low mass galaxies 
to the star formation rate at late times. 
In other words, 
not only is there less gas globally available for star formation at
the late epochs, but a smaller fraction
of this is contributed by the baryonic 
components of the smaller mass haloes than at earlier times.

So, an exact treatment of star formation in debris may be less
important for zero redshift normalisation than for intermediate
redshifts ($1\ltapprox z \ltapprox 3$).

\section{Discussion and Conclusions} \label{s-concl}

One of the ways in which bottom-up hierarchical galaxy formation
breaks the bottom-up hierarchy,
i.e. the passage of matter through a `debris' phase,
 has been studied by varying the criterion 
for merging or identifying dark matter haloes between successive
pure gravity $N$-body simulation output times, and by combining
the resulting halo merging histories with semi-analytical formulae
in order to see the effects on star formation.  

An upper bound to the loss in star formation and in stars was
modelled by using the {\em median} merging/identity criterion during
construction of the {\em halo} merging history tree,
i.e. by terminating haloes whose maximum merging percentages
were below $2\pmed-1,$ where $\pmed$ is the median merging
percentage for a given simulation and can be estimated by
eq.~(\ref{e-pmedian}). 

It was found that $\pmed$ approaches unity 
as the redshift interval for comparing haloes increases
or decreases away from $\Delta z \approx 0.4$, and that
the median distances of debris are generally {\em inside}
of the successor halo outermost radii. So, 
it is expected that the maximum possible 
dampening or truncation of star formation in matter
that passes through the debris stage can only occur (in general) 
for a temporary period over a characteristic time scale 
(redshift interval) and that the matter contributes to 
the successor halo at later times.
This is why the truncation of star formation
and loss of stellar content modelled by use of the median merging/identity
criterion provides an upper bound to possible losses. 

The maximum fraction of matter passing through a debris
stage, i.e. the minimum value of $\pmed$, was found 
to be $\pmed \approx 0.84$, i.e. about 16\% of matter is in
the debris stage for $\Delta z \sim 0.4$ and $z\sim 1$. 
(This fraction is only weakly dependent on the curvature parameters.)

The application of the median criterion was 
compared with the application of the {\em largest successor} 
merging/identity criterion, which provides a lower bound to star loss 
by linking {\em every} halo at a time step $t_i$ with its largest
successor at $t_{i+1}$, unless no particles of the halo at $t_i$ 
are present in any halo at $t_{i+1}$.

The resulting losses for star formation and luminosity functions
were found to be strongest for low luminosity galaxies and 
at intermediate redshifts $(1\ltapprox z \ltapprox 3)$. The 
losses in both cases are mostly around 0.05 to 0.10~dex, i.e.
10\%-30\%, have some dependence on the matter density parameter $\Omega_0,$
and negligible dependence on the cosmological constant $\lambda_0$
(where $h \equiv 0.65$ is kept constant).

This upper bound on likely losses in star formation rates and stellar
populations is smaller than the uncertainties in estimates of 
corresponding observational parameters, so it may not be urgent
to include a correction for this in Press-Schechter based
galaxy formation models, provided that dwarf galaxy statistics
are not of primary interest.

On the other hand, the modelling of higher order statistics such
as the shape and slope of the faint end
of the luminosity function or the evolution of galaxy 
morphology at intermediate redshifts could be expected to 
be more sensitive to the effects of debris. In modelling of 
these cases, care should be taken to cover parameter space 
which includes the redshift intervals for which maximum
debris loss occurs ($\Delta z \approx 0.4$ in the present study). 

The two choices of merging/identity criterion adopted here
bracket other cases in the literature.
Since the values of $2\pmed-1$ were found to be well above 50\%,
the $P=50\%$ criterion adopted in \citet{RPQR97} lies between 
the two cases studied. 

The modified  $P=50\%$ criterion 
adopted by \citet{KCDW99a} is slightly stricter than the 
$P=50\%$ criterion of \citet{RPQR97}, due to the requirement regarding
`merging' of the most bound particle of the progenitor halo. On the
other hand, the later `recovery' of `lost' galaxies by those authors 
weakens the criterion, although not to the extent of the largest
successor criterion, for which a single particle is sufficient to 
choose a successor halo (provided that no halo at $t_{i+1}$ contains
two particles of the progenitor halo at $t_i$). So again this is
intermediary between the cases studied here.

A useful possibility for following up the present work would be to 
study the possible {\em positive} effects of a debris stage
on star formation rates, in particular to study the possibility that
the creation of tidal dwarf galaxies could partly compensate for
the negative effects of dampened or decreased star formation rates
studied here. This would (ideally) 
require modelling of merging history trees
that break the bottom-up hierarchy by 
including the possibility of multiple successor haloes (and galaxies).
This would probably be more simple to implement in an $N$-body 
plus semi-analytical model rather than a Press-Schechter plus 
semi-analytical model.

In principle, it is possible 
that the creation of tidal dwarfs could partially 
compensate for the losses in star formation and stellar content
constrained in the present study. However, since the fraction of 
debris which can even temporarily form highly dense tidal features,
let alone permanently bound objects, seems to be no more than
around 1\% (Sect.~\ref{s-ttail}) of a halo's mass, 
even at the epochs and
for redshift intervals maximising debris loss, it is more likely 
that tidal dwarf haloes/galaxies only contribute to more subtle
statistics representing galaxy formation.

\section{Acknowledgements}
Thanks to Ariane Lan\c{c}on,  
Thanu Padmanabhan, Sunu Engineer and Jasjeet Bagla 
for helpful discussions, and the latter three for 
provision of pure gravity $N=2\e{6}$ $N$-body simulations.
  This research has been carried out with support from the 
Institut d'Astrophysique de Paris (CNRS), the Observatoire de 
Strasbourg, the 
EARA (European Association for Research in Astronomy) and the 
Polish Council for Scientific Research Grant
KBN 2 P03D 008 13.

\fflowchart

\section{Appendix: The {\arfus} freeware package}

\subsection{General description}

The freeware (GNU Public License) 
package {\arfus} is an update of the  
galaxy formation modelling method of \citet{RQP93,RPQR97}, and is intended
to be a user-friendly software package. The initial releases
are moderately user-friendly to anyone reasonably familiar with unix and
fortran and moderately familiar with concepts in 
numerical galaxy formation modelling.

{\arfus} is intended to be useful for
generating merging history trees and evolving galaxy properties,
and 
is available at {\em http://www.iap.fr/users/roukema/ArFus/index.html}. 

It combines the full non-linear
information of the collapse of dark matter haloes from (pure gravity)
$N$-body simulations with the flexibility and the rapidity of 
analytical formulae which represent gas thermodynamics and star formation
in so-called `semi-analytical' galaxy formation models 
\citep{KWG93,ColeGF94}.
{\arfus} is written primarily in Fortran 77 and should
be easily compilable on any Unix machine. (Version {\arfus}-V0.03 has
been successfully compiled and run on DEC, SUN and linux platforms,
and earlier versions also ran on IBM unix platforms.)
The structure of the
package is modularised into subroutines for the different 
logical and physical processes, and the major  
steps
(detection of haloes, creation of halo merging tree, etc.) are 
run as individual programmes. The structure is
shown schematically in Fig.~\ref{f-flowchart}. Documentation
on the subroutines is automatically extracted directly from the compilable
code and crosslinked in {\em html} format.

More details on the key subroutine {\em AFhtog}, which converts the
halo merging history tree into a combined galaxy plus halo merging history
tree and is called by {\em AFhist0} (which itself is called by 
a main programme {\em AFhist}), are provided in \SS\ref{s-afhtog}.

The choices of formulae for gas cooling and reheating (feedback) 
and star formation are focussed in special subroutines, so that
these can be easily modified by the user. Parameter values are
read in from text files, of which examples are provided in the 
package. The formulae and values used in the present study
(by way of example) are mentioned in \SS\ref{s-model} above.

Different levels of use are possible depending on the user's
interest: 
\begin{list}{(\roman{enumi})}{\usecounter{enumi}}
\item The simplest, black box, approach to the package would be to use 
output lists of galaxy absolute magnitudes and positions
at desired output redshifts 
which have already been calculated
from galaxy plus halo merging history trees derived from 
$N$-body simulations for standard cosmological parameter values
and for `reasonable' options for other parameters.
\item A less trivial use of the package would 
be to read in the pre-calculated halo plus galaxy merging trees,
but to modify the input choices of star formation formulae, and 
then to generate lists of galaxy absolute magnitudes and positions
at desired output redshifts. 
\item An example of 
more sophisticated usage would be to 
derive halo merging history trees via
{\arfus}  from the user's own gravity-only $N$-body simulations.
\end{list}

\ffhtogA

\ffhtogB

\subsection{{\em AFhtog}: conversion of a halo merging history tree
into a galaxy plus halo merging history tree}\label{s-afhtog}

The most massive halo
among a set of haloes which merge together between $t_i$ and $t_{i+1}$
to form a single halo contains a galaxy 
(unless no stars have formed), which 
is considered to be the `central'
galaxy. Its halo is the `central' halo for this system.
The other haloes initially contain galaxies, and
are labelled `satellite' haloes. After the merger of a satellite
halo with a central halo, the satellite galaxy will (in general)
take some time before it merges with the central galaxy due
to dynamical friction.

(0) (zeroth step) An initial galaxy+halo merging history tree (MHT) 
is created with one galaxy per halo
at all times, i.e. galaxy merging is identical to halo merging.
This is done by re-interpreting the variables 
representing the halo MHT
to represent both a galaxy MHT and (initially) 
the halo MHT. When in later steps halo
merging is distinguished from galaxy merging, a new 
variable (in effect, though not formally, a pointer)
is added to enable the halo MHT to be represented via
the galaxy MHT. The nodes of the halo MHT form a subset of the
nodes of the galaxy MHT.
In addition, other variables
representing useful
(but partly redundant) information on the combined galaxy+halo 
MHT are added.

The only subsequent modification to this is that when a satellite
halo merges with a central halo, in general the galaxy
which was in the satellite halo will not merge immediately
with the central galaxy, so a `new' value of the galaxy/halo 
index 
i.e.
a `new haloless' galaxy, is created
to represent the satellite galaxy at each
subsequent output time until the merger occurs.
Physically, the satellite galaxy is orbiting within the
central halo, but this link is represented indirectly.
Since the galaxy/halo index of this galaxy only represents a 
galaxy, i.e. does not represent any halo during the period 
from $t_i$ to $t_{i+1}$, the galaxy is termed
`halo-less'. This is just an algorithmical convention, 
which distinguishes satellites which merge `fast' from
those which merge `slowly' [see (c) and (d) below for
definitions of `fast' and `slow'].

The following steps, illustrated schematically in
Figs~\ref{f-f_htogA} and \ref{f-f_htogB} are then carried out.

(1) Halo merging times are {\em interpolated} between time steps,
in order that halo merging does not just happen (in general)
in multiple mergers at the (N-body simulation) output time steps.
(See \SS2.2.5 and fig.~14 of \citet{RPQR97} for more discussion.)
An option is provided to have multiple mergers at the  
{\em midpoints} of the time interval instead of using
the recommended isothermal potential interpolation.

(2) A loop through the output times from earliest to latest
is carried out.
At each output time, the following steps are carried out.

(a) A dynamical friction type formula is used to 
calculate how long, in principle, a satellite
galaxy would take to merge with the central galaxy. This
calculation is independent of the halo merging time calculated 
at step (1),
for simplicity, except that the halo merging time is used
to calculate the cosmological epoch. In the simplest case
this time lapse is labelled  $t_g^1$.

(b) The calculated halo merger time $t_h$ and
$t_g^1$  are used to estimate when the satellite galaxy
should, in principle, merge with the central galaxy.
(Note that both are time {\em intervals}, not cosmological times.)

(c) If a satellite would merge with the central galaxy
before time $t_{i+1}$, then the linked parts of the 
MHT structure, which represent merging/identity between
output times, do not need to be modified at all for
that satellite galaxy and its halo. The fact that both
the satellite halo and galaxy merge with the central
halo and galaxy (in that order) during the interval
between  $t_i$  and  $t_{i+1}$ is represented by 
$0< t_h < t_h+t_g^1 < t_{i+1} - t_i$.

This case is referred to as `fast' galaxy merging in 
Fig.~\ref{f-f_htogA}.

(d) If a satellite would merge with the central galaxy
{\em after} time  $t_{i+1}$, then the linked parts of the 
MHT structure require modification, possibly several
output times later. This is the first cause of the
complexity of {\em AFhist0}. 

This case is referred to as `slow' galaxy merging in 
Fig.~\ref{f-f_htogA}. The difficulty is that a new `halo-less'
galaxy needs to be added, which may take several time
steps before finally merging with the central galaxy.

(e) The second cause of complexity  
is that in the same slow merging 
case (d) with a satellite galaxy  $S_1$
and a central galaxy  $C_1$, the central galaxy may 
itself become a satellite galaxy  $C_1=S_2$ 
and merge with its
own central galaxy  $C_2$  at or before the time when the
original satellite ($S_1$) is calculated to merge. Because
the dynamical friction type formula cannot be as good as
(or identical with) the N-body simulation dynamics,
it is possible that  $S_1$  merges with  $C_1=S_2$  {\em after}
$C_1=S_2$ merges with  $C_2$. Physically, while this is not
totally impossible, it seems more reasonable, and is
easier computationally, to decide that  $S_1$  merges  
with  $C_2$  at the time when  $C_1=S_2$  merges with  $C_2$.
This is the option chosen here. That is, during
the interval  $(t_{i+j}$, $t_{i+j+1})$  when  $C_1=S_2$ merges with $C_2$,
(i) the central galaxy of $S_1$ is redefined to be  $C_2$  instead
of  $C_1=S_2$, and (ii) both $C_1=S_2$ and $S_1$ merge with $C_2$
at the same point of time during that time interval.

This case is shown schematically in Fig.~\ref{f-f_htogB}.
Although this case is rare, it needs to be treated algorithmically
in one way or another if case (d) is allowed and if the galaxy MHT
is to be logically complete. 

If case (d) is not allowed, i.e. if all 
galaxy mergers are forced to occur before the output time $t_{i+1},$
then unless a very large $N$-body simulation output time step is
chosen, the galaxy mergers would follow the halo mergers very quickly, 
so that little difference between the halo and galaxy MHT's would be
possible.

If the galaxy MHT is not logically complete, then errors in the evaluation
of the star formation history tree, such as reading or writing invalid
bytes in memory would be possible.

Hence, inclusion of both cases (2d) and (2e) as outlined here seems
necessary.


\end{document}